\pdfoutput=1
\documentclass{article}
\usepackage{graphicx} 

\usepackage[preprint]{custom}

\usepackage[authoryear,round]{natbib}  

\usepackage[utf8]{inputenc} 
\usepackage[T1]{fontenc}    
\usepackage{hyperref}       
\usepackage{url}            
\usepackage{booktabs}       
\usepackage{amsfonts}       
\usepackage{nicefrac}       
\usepackage[protrusion=true,expansion=false]{microtype}      
\usepackage{xcolor}         
\usepackage{subcaption}

\title{Overcoming Vision Language Model Challenges in Diagram Understanding: A Proof-of-Concept with XML-Driven Large Language Models Solutions}

\author{%
Shue Shiinoki \quad  Ryo Koshihara \quad  Hayato Motegi \quad  Masumi Morishige\\ \\
Galirage Inc. \\ info@galirage.com
}

\begin{document}

\maketitle
\renewcommand{\thefootnote}{\fnsymbol{footnote}}

\begin{abstract}
    Diagrams play a crucial role in visually conveying complex relationships and processes within business documentation. Despite recent advances in Vision-Language Models (VLMs) for various image understanding tasks, accurately identifying and extracting the structures and relationships depicted in diagrams continues to pose significant challenges. This study addresses these challenges by proposing a text-driven approach that bypasses reliance on VLMs’ visual recognition capabilities. Instead, it utilizes the editable source files—such as xlsx, pptx or docx—where diagram elements (e.g., shapes, lines, annotations) are preserved as textual metadata.  In our proof-of-concept, we extracted diagram information from xlsx-based system design documents and transformed the extracted shape data into textual input for Large Language Models (LLMs). This approach allowed the LLM to analyze relationships and generate responses to business-oriented questions without the bottleneck of image-based processing. Experimental comparisons with a VLM-based method demonstrated that the proposed text-driven framework yielded more accurate answers for questions requiring detailed comprehension of diagram structures.The results obtained in this study are not limited to the tested .xlsx files but can also be extended to diagrams in other documents with source files, such as Office pptx and docx formats. These findings highlight the feasibility of circumventing VLM constraints through direct textual extraction from original source files. By enabling robust diagram understanding through LLMs, our method offers a promising path toward enhanced workflow efficiency and information analysis in real-world business scenarios.
    The related code is available at \url{https://github.com/galirage/spreadsheet-intelligence}, which provides the core library developed for this research. The experimental code using this library can be found at \url{https://github.com/galirage/XMLDriven-Diagram-Understanding}.
\end{abstract}

\section{Introduction}
\label{sec:intro}
Diagrams, which visually organize and present information, are one of the most powerful methods for combining text and visual representation in documents. For instance, diagrams are widely utilized in business documentation as effective tools for information communication, such as explicitly illustrating relationships between elements in system architecture blueprints or clarifying complex procedures and processes step by step using flowcharts. 
Recently, with the rapid development of large language models (LLMs)\citep{openai_gpt-4_2024, anil_palm_2023}, methods for processing textual information have become increasingly sophisticated. Applications such as QA systems combined with knowledge bases derived from documents \citep{krishna_fact_2024, fleischer_rag_2024}, efficient document creation and information sharing through automated summarization and report generation \citep{thoppilan_lamda_2022, liu_mmbench_2025}, and automation of compliance verification tasks \citep{cava_safeguarding_2024, kande_security_2024, sollenberger_llm4vv_2024} have made it possible to automate complex tasks traditionally performed manually. This has significantly accelerated the efficiency and standardization of business processes.

\begin{figure}[tbp]
    \vspace{-5pt}
    \centering
    \includegraphics[width=0.85\textwidth]{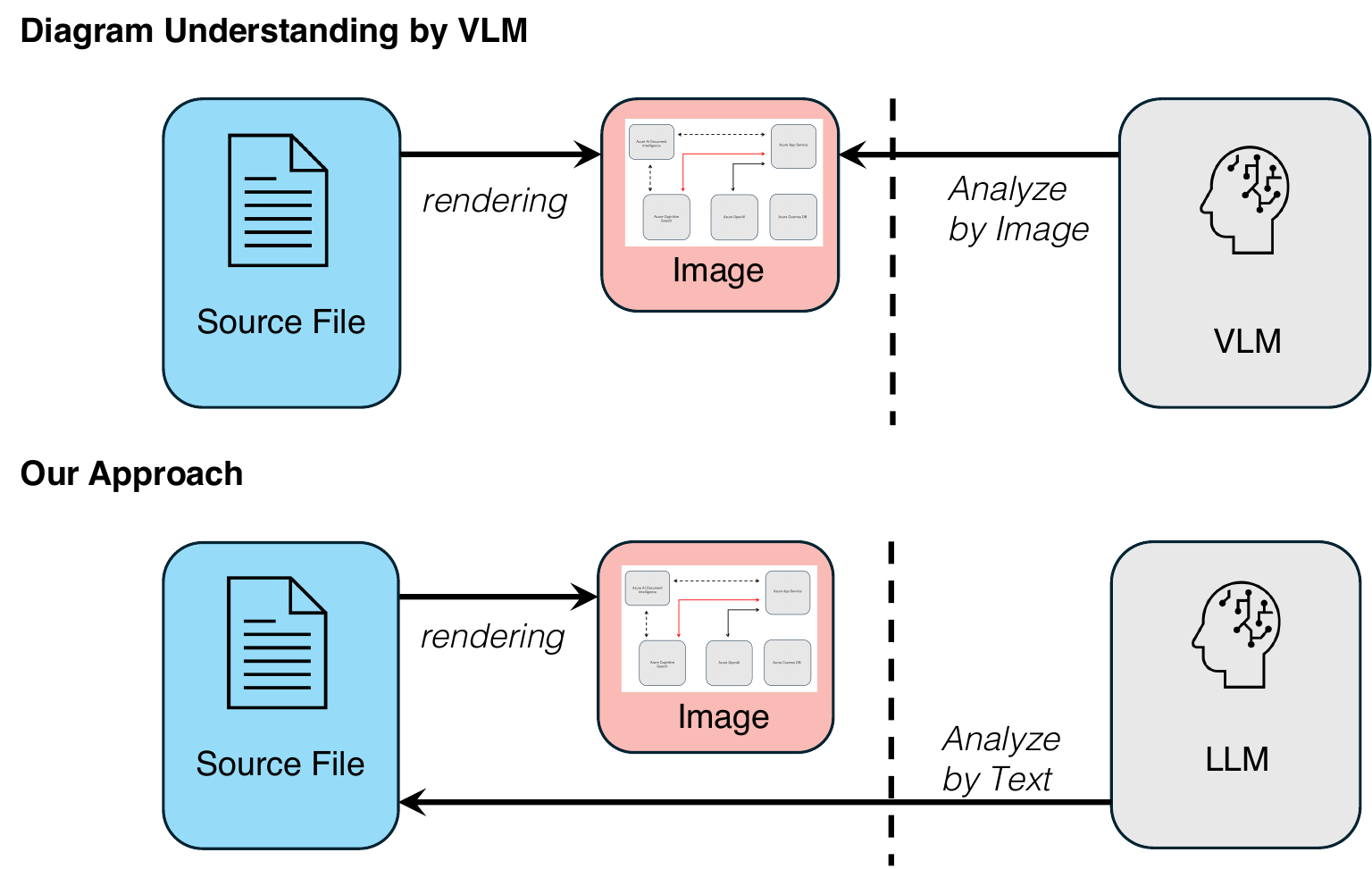}
    \caption{\small
    Our approach does not rely on the visual recognition of the VLM, but rather inputs and analyzes the LLM as text, with the underlying graphic information stored in the source files referenced to render the Diagram image.
    }
    \label{fig:concept}
    \vspace{-10pt}
\end{figure}

Moreover, with the advent of Vision-Language Models (VLMs) capable of understanding images \citep{openai_gpt-4o_2024, noauthor_gpt-4vision_2023, fu_vita-15_2025, noauthor_introducing_2024}, there is increasing potential to enhance workflow efficiency by processing not only textual information but also diagrams as images.
VLMs, particularly proprietary models, have achieved remarkable results in diverse tasks such as generating captions for general images like photographs and illustrations, and visual question answering (VQA) \citep{openai_gpt-4o_2024, noauthor_introducing_2024}
Additionally, significant progress has been made in visual reasoning tasks that analyze more symbolic visual information, such as diagrams, fostering growing expectations for automated systems that integratively analyze text and diagrams\citep{lu_mathvista_2024, zhang_mathverse_2024, chen_how_2024}

While these new possibilities are expanding, it is not easy to accurately grasp and extract the structure and interrelationships among the elements shown in a diagram, and many challenges still remain to be overcome before they can be implemented in actual work. 
For example, VLM has been shown to be limited in geometric recognition of images composed of shapes \citep{kamoi_visonlyqa_2024, zhang_mathverse_2024, yue_mmmu_2024}, and the stage of correctly textualizing geometric structures is reported to be a bottleneck in question-answering (QA) and inference tasks using diagrams \citep{ye_beyond_2024}.
In tasks such as question answering (QA) or reasoning about diagrams, the stage of correctly textualizing geometric structures has been reported as a bottleneck \citep{ye_beyond_2024}.
In addition, images containing complex geometric primitives, such as decorative representations of overlapping adjacent lines and shapes, are prone to misidentification by VLMs \citep{rahmanzadehgervi_vision_2024}, and existing VLMs are not robust enough for the variety of color schemes and drawing formats found in real documents \citep{ye_beyond_2024, singh_flowvqa_2024, tannert_flowchartqa_2023}.
Furthermore, while VLMs are relatively good at recognizing entities (graphics) such as system blueprints and flowcharts, it is still difficult to accurately grasp the relationships between elements represented by lines\citep{giledereli_vision-language_2024}, and they tend to halucinate when their previously learned knowledge is inconsistent with the input visual information\citep{mukhopadhyay_unraveling_2024, giledereli_vision-language_2024}.
Considering these limitations, it is essential to develop methods that mitigate the visual capability constraints of VLMs to effectively utilize language models in business scenarios and other contexts involving visually enriched documents.

In this study, we propose a novel approach that does not process diagrams as images but instead extracts and transforms the shape information stored within source files, enabling analysis of this information as text data using 'LLMs'. In business scenarios, documents often exist in editable source file formats such as docx, xlsx, and pptx before being converted into PDF format \citep{noauthor_isoiec_nodate}.  The entity of this source file is XML, which can be parsed to directly extract information on shapes, lines, annotations, and other elements that make up the diagram, and this information can be provided to LLM as text input to bypass visual recognition bottlenecks and enable comprehensive understanding and analysis of diagrams. In this study, system design documents from requirements definitions created in Excel, commonly used in Japanese system development workflows, were prepared. Using these documents, we conducted evaluations by posing several business-relevant questions to compare the proposed method with a VLM-based approach. The results confirmed that while VLMs were unable to answer certain questions accurately, the proposed method provided accurate answers. This study demonstrates the potential of analysis through source file, such as XML, processing to overcome the challenges of diagram comprehension in VLMs. It highlights the possibility of improving workflows and achieving greater efficiency in information utilization systems leveraging LLMs in business contexts.

\section{Related Works}
\subsection{Visual Recognition Limitation of VLM in Diagram Understanding}
As various verification studies have been reported in the development of VLM, challenges in its visual recognition performance have come to light.
For example, VLM has been shown to be limited in geometric recognition of images composed of shapes \cite{kamoi_visonlyqa_2024, zhang_mathverse_2024, yue_mmmu_2024}, and the stage of correctly textualizing geometric structures is reported to be a bottleneck in question-answering (QA) and inference tasks using diagrams \cite{ye_beyond_2024}.
In tasks such as question answering (QA) or reasoning about diagrams, the stage of correctly textualizing geometric structures has been reported as a bottleneck \cite{ye_beyond_2024}.
In addition, images containing complex geometric primitives, such as decorative representations of overlapping adjacent lines and shapes, are prone to misidentification by VLMs \cite{rahmanzadehgervi_vision_2024}, and existing VLMs are not robust enough for the variety of color schemes and drawing formats found in real documents \cite{ye_beyond_2024, singh_flowvqa_2024, tannert_flowchartqa_2023}.
Furthermore, while VLMs are relatively good at recognizing entities (graphics) such as system blueprints and flowcharts, it is still difficult to accurately grasp the relationships between elements represented by lines\cite{giledereli_vision-language_2024}, and they tend to halucinate when their previously learned knowledge is inconsistent with the input visual information\cite{mukhopadhyay_unraveling_2024, giledereli_vision-language_2024}.
Considering these limitations, it is essential to develop methods that mitigate the visual capability constraints of VLMs, such as the method presented in this study, to effectively utilize language models in business scenarios and other contexts involving visually enriched documents.

\subsection{The Role of Source File Information in Improving Visual Recognition and Understanding by VLMs and LLMs}
When enabling VLMs and LLMs to understand rendered formats of source files that humans typically view, there are several examples of using source file information as input to facilitate their understanding. 
In the field of VLM-based GUI understanding, while screenshots of screens are input as images, the performance of recognition is enhanced by simultaneously providing layout and positional information of widgets and buttons, along with textual descriptions, as text input \cite{you_ferret-ui_2024, li2024ferret}. Moreover, methods that directly input sources containing metadata such as text information and document structure, like HTML or docx, into models have been reported to improve the accuracy of document comprehension and analysis tasks in VLMs \cite{xu2020layoutlmv2, huang2022layoutlmv3, tan2024htmlrag}.
In the context of diagram comprehension, while there are no studies explicitly demonstrating the advantages of using metadata information from source files, many results suggest its potential. For example, a study analyzing the task of flowchart comprehension, which separates the phase of generating Mermaid notation from flowchart images through visual recognition and the phase of answering questions about the flowchart using the Mermaid text representation, demonstrated that the bottleneck lies in the image-to-Mermaid conversion\cite{ye_beyond_2024, pan_flowlearn_2024}. It also showed that once the correct topology is understood, answering questions about the flowchart is not challenging. These findings, combined with the multiple reports of the visual recognition limitations of VLMs highlighted in the previous section, support the effectiveness of utilizing source file information in diagram comprehension within documents and align with the results of this study.

\section{Methodology}
\label{sec:methodology}
\subsection{Diagram Information Parsing from Xlsx}
\begin{figure}
    \includegraphics[width=\textwidth]{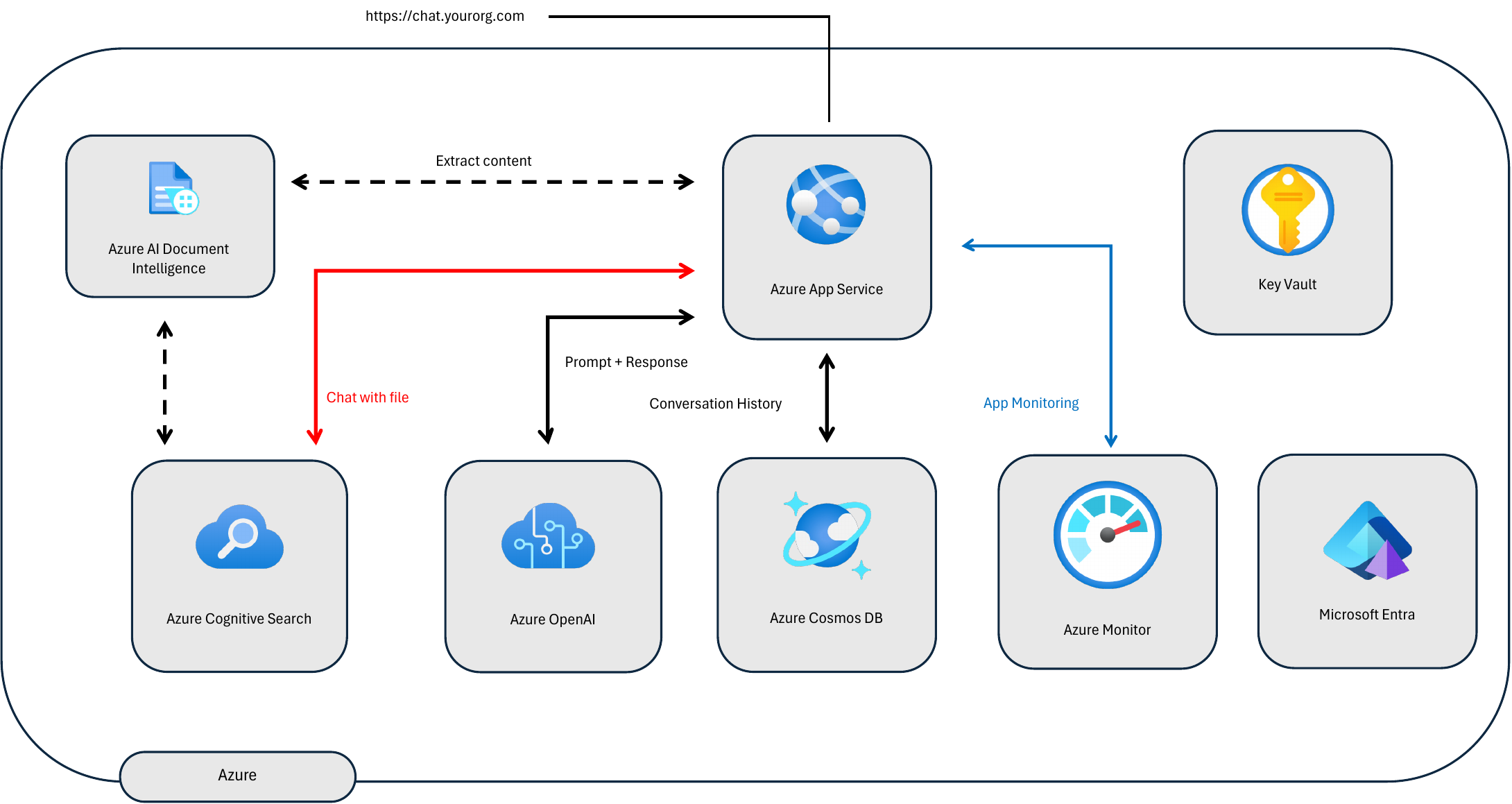}
    \caption{The diagram used in this study, a system blueprint drawn in Excel using rectangles, text boxes, straight connectors, and bent connectors.}
    \label{fig:arch}
\end{figure}
\begin{figure}
    \includegraphics[width=\textwidth]{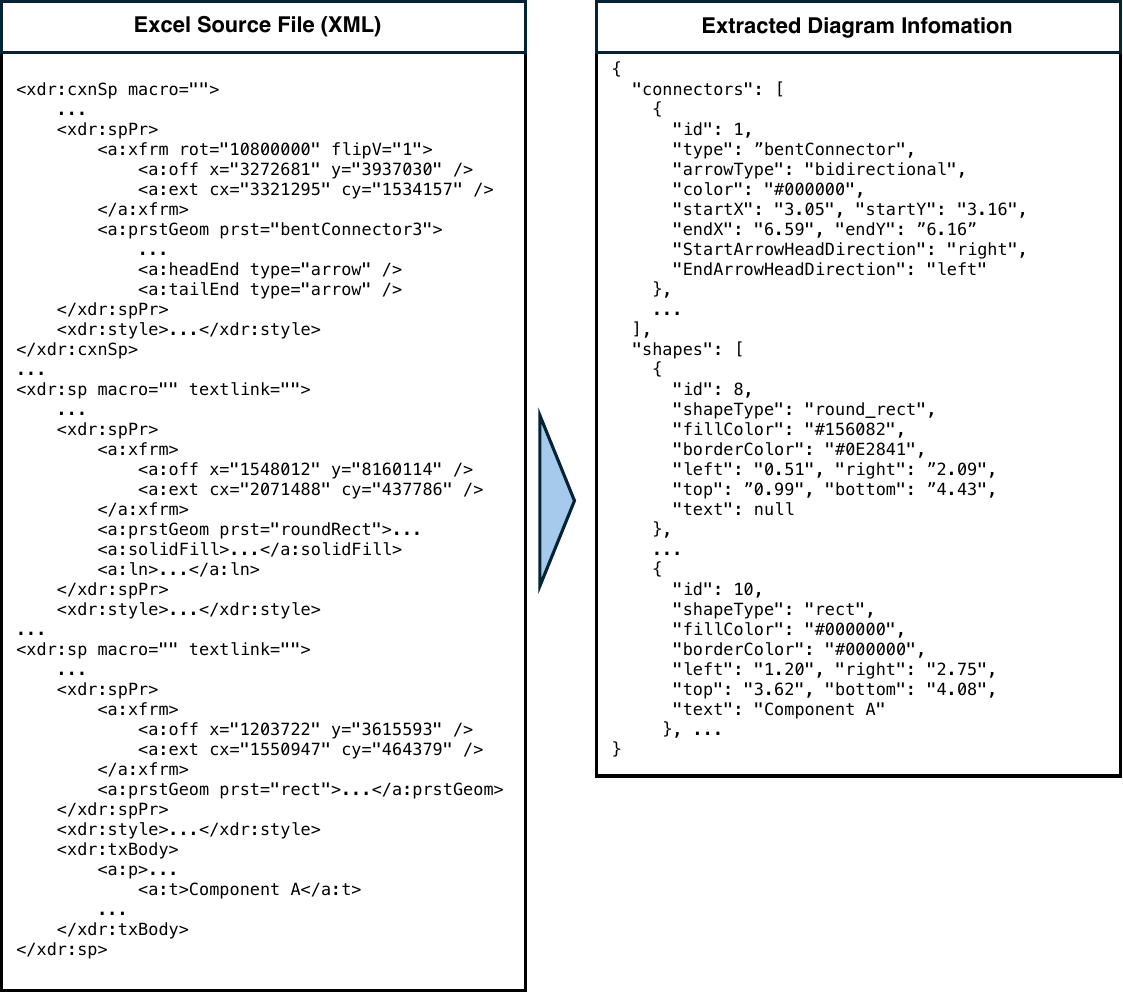}
    \caption{\small
    On the left is an XML excerpt of shape data from an xlsx file, difficult to interpret due to raw numerical values and aliased strings. It also contains excessive, scattered information across multiple files. On the right is a JSON format that parses, transforms, and summarizes meaningful information on connectors and shapes, optimized for diagram understanding and LLM input.
    }
    \label{fig:data_trans}
\end{figure}
As mentioned earlier, this study proposes a workflow that allows LLMs to understand and analyze diagrams and answer related questions by extracting and converting diagram information stored in source files, without treating documents as visual information. The extracted information is appropriately formatted and converted into text.

First, since the information constituting the document is stored in XML format within the source file, we developed a library to parse this XML and extract diagram-related information. In this study, we targeted system design diagrams saved as .xlsx files created using Microsoft Excel (Fig. \ref{fig:arch}). Documents created with Office software are globally prevalent and enabling LLMs to analyze these documents is highly desirable. However, while data loaders such as LangChain Unstructured and Azure Document Intelligence are available, these loaders are primarily limited to loading textual data and do not parse visual elements into text.

Moreover, in Japan, there is a significant reliance on Excel for creating requirement definition documents and system design diagrams. Therefore, there exists a substantial demand to extract not only table data but also diagrams from Excel. System design diagrams consist of shapes, lines, and text that explain the concepts represented by these shapes and lines. These diagrams are a representative form of diagrams where shapes denote entities and lines represent relationships. For these reasons, this study specifically targeted system design diagrams saved in .xlsx files.

To achieve this, we developed a library that parses source files in .xlsx format and extracts diagram-related shape information (\url{https://github.com/galirage/spreadsheet-intelligence}). The data extracted using this library is shown in Fig. \ref{fig:data_trans}. The source files of Office documents, including .xlsx, are essentially XML files compressed into a zip format. By decompressing and reading the XML, it becomes possible to retrieve the elements that constitute the document. The XML contains information about user-placed objects such as text, shapes, and text boxes, along with editable attributes like drawing position, rotation, theme, color, shape formatting, and text formatting.

An example of a system design diagram created in Excel is shown in Fig. \ref{fig:arch}. This diagram consists of objects such as text boxes, rectangles, straight connectors, and bent connectors. We extracted information related to these four types of objects as the components of the diagrams.

\subsection{Parsing and Transforming Diagram Elements for Enhanced Analysis}
Subsequently, the extracted shape information was transformed into a format that the LLM could easily understand. 
Information extracted from XML cannot deliver sufficient analysis accuracy if used as-is. Therefore, we chose to extract only information relevant to the diagram for rectangles, text boxes, straight connectors, and bent connectors.
For shapes and connectors, coordinate information was first extracted. The coordinate information for shapes and connectors, as provided in the XML, does not reflect rotation, flipping, scaling, or other transformations. To address this, a conversion process was applied to incorporate all these factors, ensuring the values accurately represent the appearance of the rendered diagram.
Additionally, for fill, border, and line color information, cases where the color was specified using software-specific color theme names or where the XML data did not correspond directly to the rendered state were also handled through conversion.
On diagrams, relationships are expressed by connecting shapes representing entities with connectors. While the connection information between connectors and shapes is saved if the diagram creator explicitly configures it during editing, diagrams can also represent relationships with disconnected shapes and connectors. In such cases, it is impossible to determine the connections between shapes and connectors directly from the XML. Therefore, instead of relying on the source file's connection information, we opted to decode the direction of each connector's endpoints into four cardinal directions based on the connector's coordinate data. This enables understanding which direction a connector is pointing and identifies the shapes connected to each connector.
Furthermore, Office software shapes can be grouped to represent collections of shapes. However, since the way these groups are created varies by the diagram creator, group information was not utilized as input data in this process.
The parsed information was then organized, as shown in Fig. XX, into JSON format with attributes grouped for each shape (including both text boxes and rectangles) and each connector (including both straight and bent connectors).

\section{Results}
\label{sec:result}

\subsection{Entity Understanding Using the XML-driven Approach}
\label{sec:result1}
\begin{figure}[tbp]
    \centering
    \includegraphics[width=\textwidth]{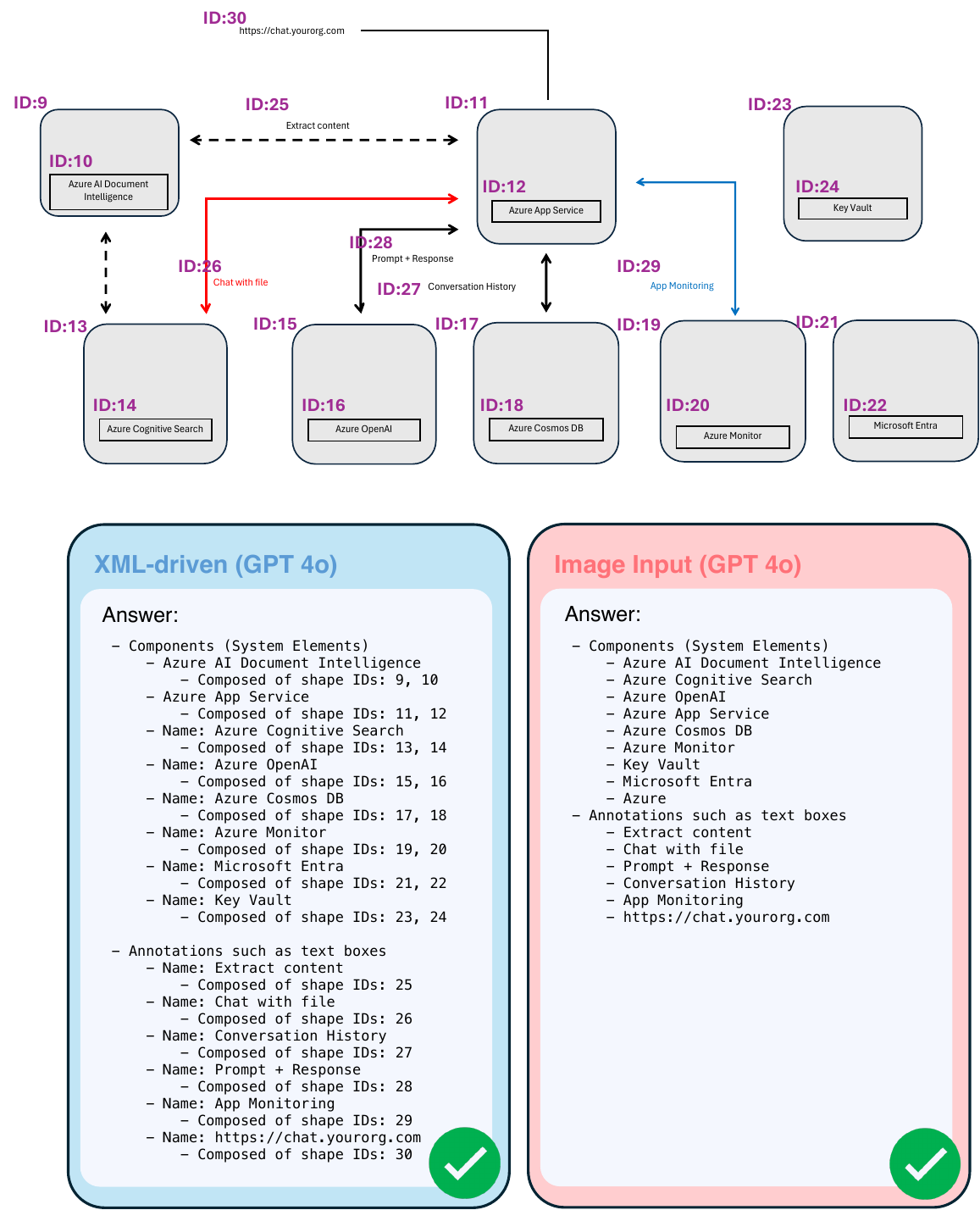}
    \caption{\small
    Comparison of the output of the proposed method (Text-driven) and VLM (Image Input) for understanding entities in a diagram.
The figure above shows the input system design diagram. The ID assigned to each shape indicates the ID assigned by the proposed method when passing graphic information by text. Entity comprehension is correct for both methods without omission, and the results show that the XML-driven approach is able to understand components consisting of text boxes and round-rectangles without any group structure information.
    }
    \label{fig:shape_cs}
\end{figure}

We present an example of analyzing diagrams with diagram information extracted from XML to examine how they contribute to understanding system design diagrams. First, we focus on understanding each entity, such as components and text boxes, in the system design diagram shown in Fig. \ref{fig:arch}. This includes components composed of rectangles and text boxes, annotation information for connectors using text boxes, and text boxes representing entry points, such as URLs. Using a diagram converted into JSON format based on the previous method, along with prompts that describe each attribute, we instructed the system to categorize and list components, annotations, and other text boxes. Note that icon images were excluded from the input in this instance. The JSON format contains information such as the positions of the four edges of the rectangles and text boxes (“left,” “right,” “bottom,” “top”). For text boxes, the strings contained within the boxes are also recorded. Additionally, attributes such as fill color, border color, and whether the shape is a rounded rectangle or a plain rectangle were provided. Fig. \ref{fig:shape_cs} shows the results obtained by instructing the LLM to understand and enumerate the entities in the diagram. As a result, all components and annotations were correctly enumerated, and the text boxes for annotations were also accurately listed. Furthermore, since no grouping information was included in this input data, pairs of text boxes representing component names and the surrounding rectangles had to be determined using the positional and other information contained in the JSON. When the IDs of shapes constituting each component were also output, it was confirmed that the system correctly identified and grouped the relevant shapes.

\subsection{Relation Understanding Using the XML-driven Approach}
\begin{figure}[tbp]
    \centering
    \includegraphics[width=\textwidth]{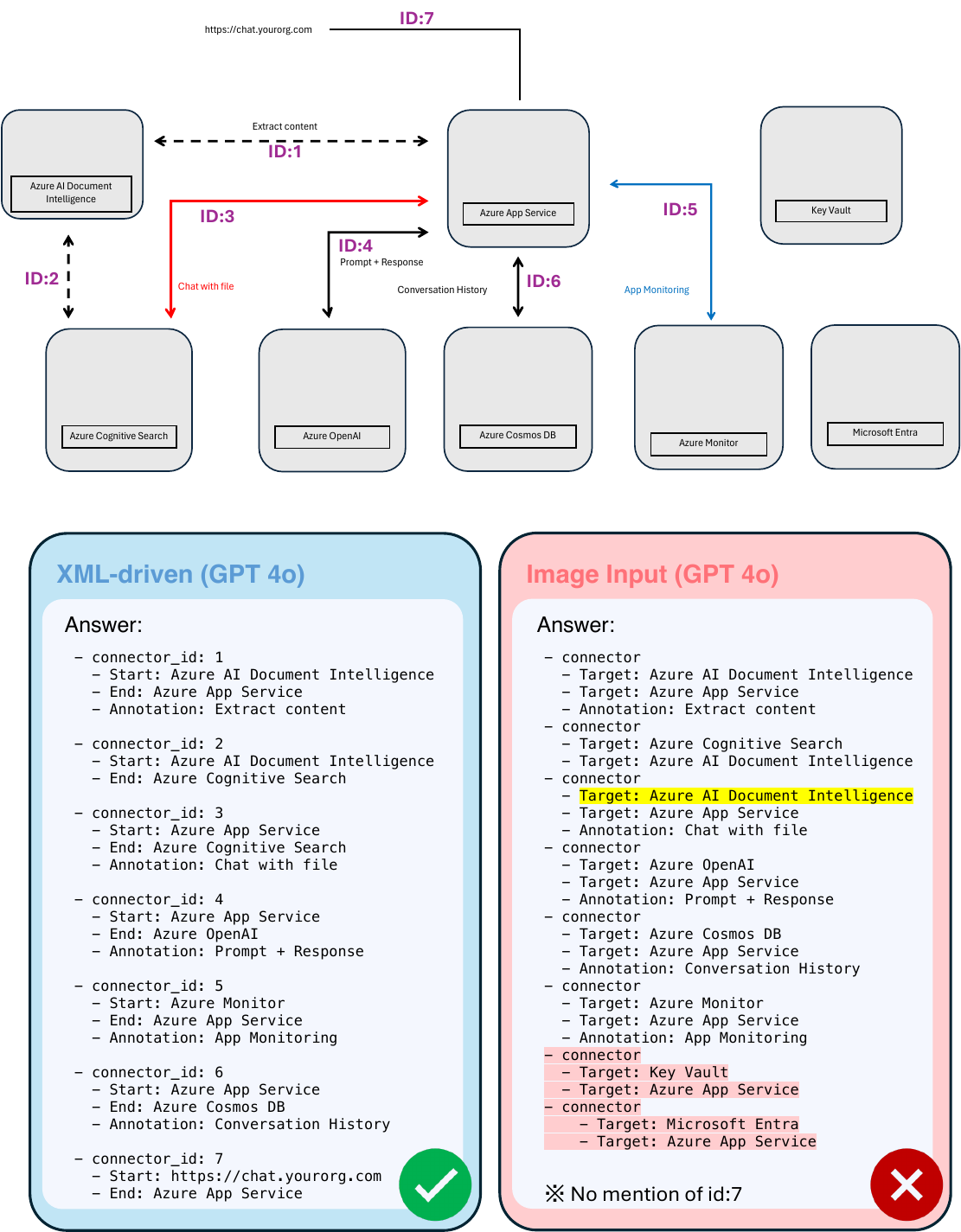}
    \caption{\small
    Output comparison of the proposed method (XML-driven) and VLM (Image Input) for understanding relationships in a diagram.
The figure above shows the ID assigned to each connector. The connector understanding is correct without omission in the XML-driven approach, while in the Image Input, yellow: probably did not detect the connector bend, red: misidentified a relationship that does not exist, and detected a connector with ID:7 where one of the objects is a text box. The image input is yellow: probably not detecting a connector bend.}
    \label{fig:connector_cs}
\end{figure}

Next, we examined whether the system could correctly analyze information about connectors that represent relationships between components. This example includes connectors with different shapes, such as straight-line connectors and curved zigzag connectors. In this case, we included both the output of the previously analyzed shape information (components and annotation details) and the JSON-formatted diagram as prompts. We instructed the system to output all connectors, specifying the two target components they connect and any associated annotations if applicable. The analysis results for the shape information included the names of each component and the positional information about the areas where the components are located. The same applies to the information on annotation text boxes. The JSON-formatted information for the connectors includes the (x, y) coordinates of the start and end points for both straight and zigzag connectors, as well as the directions indicated by each endpoint. Using this information, we instructed GPT-4 to analyze the data. Fig. \ref{fig:connector_cs} shows the results. As a result, the system correctly identified and extracted the targets for all connectors. Connectors were correctly detected for bent connectors, straight connectors, and even for connectors that were not component-to-component connectors, such as ID:7, which connects a text box to a component.
Also, the “Key Vault”, “Microsoft Entra”, and other areas where connectors do not exist were answered without any halcination.

\subsection{Comparision with VLM}
\label{sec:result2}
Using GPT-4o’s image input, the same system design diagram was analyzed.
First, a task was conducted to list all components and annotations from the image input.  Fig. \ref{fig:shape_cs} shows the results.  As a result, all component names were comprehensively listed without omissions. Similarly, annotation texts were also fully captured without any missing information. Next, we tested its ability to analyze relations, giving instructions similar to those used in our proposed method.  Fig. \ref{fig:connector_cs} shows the results. While straight connectors were correctly identified as linking two targets, hallucinations were observed for curved connectors. For example, a curved connector that actually linked Azure Cognitive Search and Azure App Service was misidentified as connecting to Azure AI Document Intelligence, which happened to be near the corner of the curve. Additionally, a connector between Azure App Service and the URL text box was missing from the output. Furthermore, connectors were mistakenly recognized between Azure App Service and unrelated components such as "Key Vault" and "Microsoft Entra", where no connectors actually existed.
In comparison with our proposed XML-driven analysis, processing diagram information extracted directly from source files—without relying on visual recognition—proved successful in enabling LLMs to understand the diagram without being affected by the detection errors inherent in visual recognition.

\section{Conclusion}
In this paper, we propose the idea of constructing a diagram-understanding method that uses an LLM driven by information extracted from source files, such as XML of Excel, rather than relying on visual recognition. Previous, image-driven approaches utilizing VLMs exhibit insufficient visual capabilities, as demonstrated in the hallucination examples in this study, indicating that there are still challenges for real-world applications such as in business contexts. In response, we presented a XML-driven diagram-understanding solution—one that leverages shape data extracted from the original source files used to create the diagrams—and highlighted its practical potential.

However, since our study is limited to preliminary experiments with a specific system design diagram, we have not thoroughly examined statistical rigor or the generalizability of our evaluation data. To validate the effectiveness of this method on a broader scale, further testing on large and diverse datasets will be necessary. Nevertheless, the library we developed to parse the XML source files of Excel documents used in this study is publicly available as open-source software, ensuring reproducibility and applicability. This enables the research community to replicate our method in diverse scenarios, refine it, and further advance its development.

Although there are constraints due to the lack of fully rigorous statistical procedures and insufficient guarantees of large-scale data generalizability, we anticipate that open-sourcing our library will stimulate further collaborative research and industrial applications. Our hope is that this work will spur the practical application of diagram-understanding technologies using language models, boost the accuracy of requirement definitions and system design analysis in business contexts, and ultimately contribute to improved operational efficiency in Japan’s system development landscape.


\newpage

\appendix
\section*{Appendix}
\renewcommand{\thesubsection}{A.\arabic{subsection}}

\subsection{Limitations}
\begin{itemize}
\item \textbf{Potential XML parsing issues due to Excel version differences:} Depending on the Excel version or individual settings, the underlying XML structure and attributes can vary, possibly leading to parsing inconsistencies or errors.
\item \textbf{Limited support for diverse objects within Office files:} Our approach currently targets standard shapes and connectors. Grouped or composite shapes, SmartArt, and specialized WordArt objects may not be fully supported.
\item \textbf{JSON input format constraint:} While we use JSON to provide shape and connector information to the LLM, alternative structured formats or more natural-language-like representations could be explored for improved flexibility or readability.
\item \textbf{Insufficient quantitative evaluation:} The effectiveness of the proposed approach has been illustrated primarily through case-based examples. Objective metrics such as accuracy or F1-scores would better validate performance.
\end{itemize}

\subsection{Future Work}
\begin{itemize}
\item \textbf{Applying the method to other Office file formats:} Extending this approach to PowerPoint (.pptx), Word (.docx), or Visio (.vsdx) opens up broader use cases and ensures generalizability.
\item \textbf{Handling more complex, real-world requirement documents:} Business documents often contain intricate layouts and multiple layers of grouping. Verifying how well our method can cope with these complexities is crucial for industrial adoption.
\item \textbf{Establishing XML-oriented design guidelines:} A potential methodology is to standardize document creation rules—avoiding excessive grouping or overly complex objects—to streamline both human readability and machine parsing in real business processes.
\item \textbf{Exploring RAG (Retrieval-Augmented Generation) approaches:} When using Excel or other Office files as data sources, it is important to investigate how to integrate RAG efficiently and manage data for effective retrieval.
\item \textbf{Hybridizing text- and vision-based methods:} In certain scenarios, combining VLM-based visual recognition (e.g., for icons) with XML-based extraction of connectors and shapes may offer a robust, hybrid solution that leverages the strengths of both approaches.
\end{itemize}

\begin{figure}[tbp]
    \vspace{-5pt}
    \centering
    \includegraphics[width=0.85\textwidth]{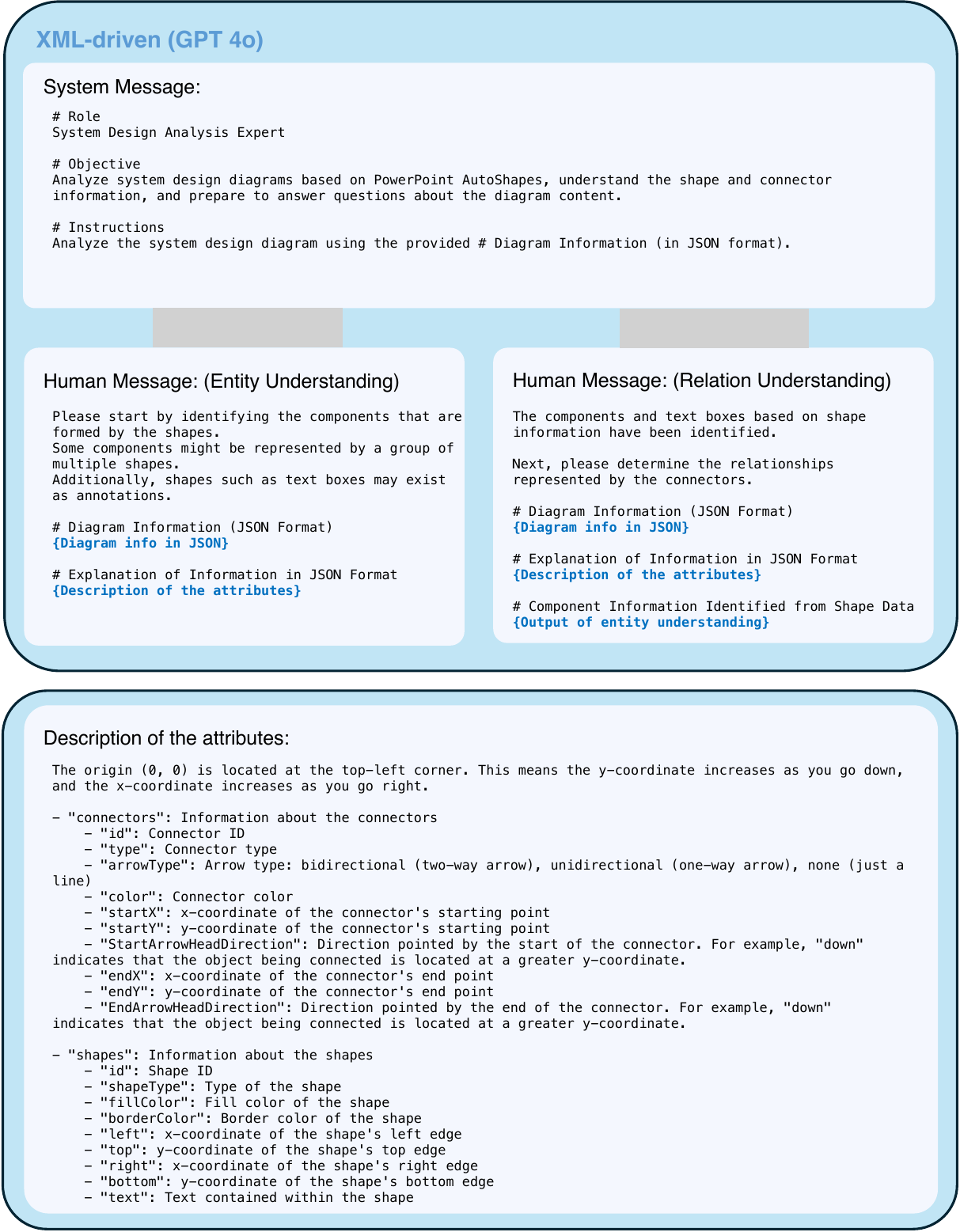}
    \caption{\small
    Prompts used in text-driven diagram comprehension. The prompts were used where the blue text was replaced with the corresponding information.}
    \label{fig:prompt1}
\end{figure}

\begin{figure}[tbp]
    \vspace{-5pt}
    \centering
    \includegraphics[width=0.85\textwidth]{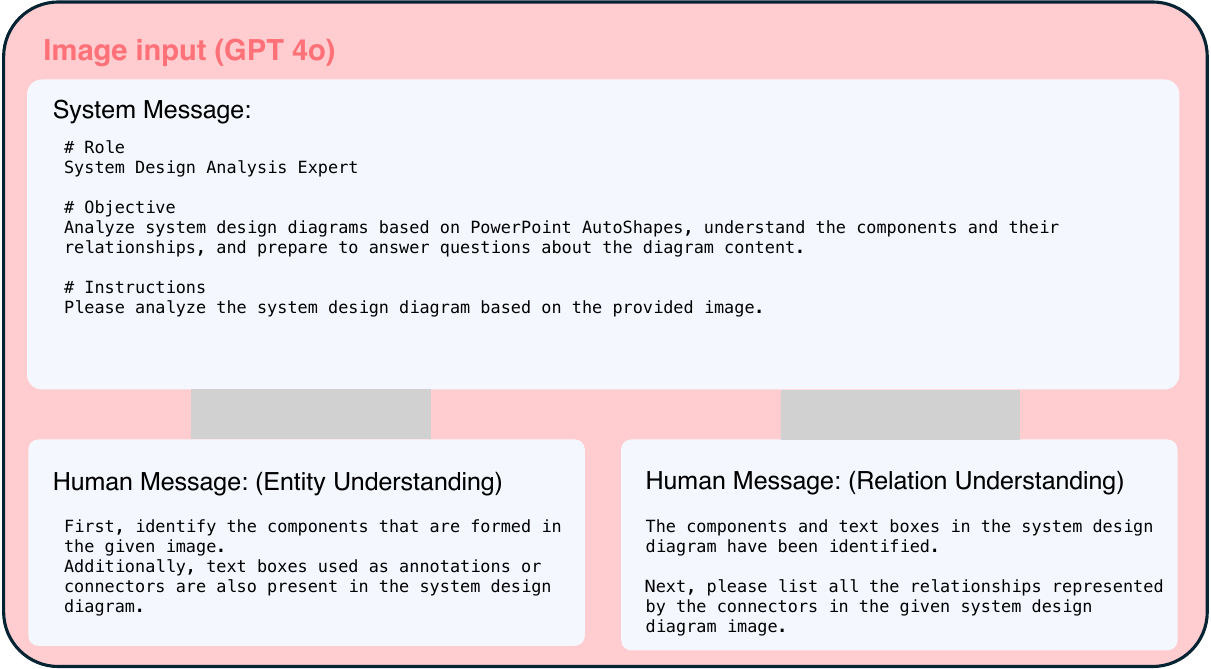}
    \caption{\small Prompts used in diagram comprehension with image input}
    \label{fig:prompt2}
    \vspace{-10pt}
\end{figure}

\subsection{Prompts Used in the Experiment}

All experiments in this study were conducted on GPT4o \citep{openai_gpt-4o_2024} (gpt-4o-2024-08-06).
The prompts used in result \ref{sec:result1} are shown in Figure \ref{fig:prompt1} and the prompts used during result \ref{sec:result2} are shown in Figure \ref{fig:prompt2}. For understanding entities on the conventions, the diagram information in JSON format shown in Figure \ref{fig:data_trans} was included in the prompts, and the information for each attribute in the JSON was given an explanation as shown in Figure \ref{sec:result1}. In addition to the JSON diagram information and information on each attribute, information on entities obtained through entity understanding was also added to the prompts to infer the relationships on the diagrams. For the image input, we used prompts that were not changed from those of the proposed method as much as possible, but only the descriptions added to give information in text were excluded.

\end{document}